\documentclass{aa}  
\usepackage{graphicx}
\usepackage{txfonts}
\usepackage{blindtext}
\usepackage{hyperref}
\usepackage{placeins}

\begin{document} 

%\title{Simulations of galaxy clusters and filaments of the cosmic web with the SKA}
\title{Filaments in and between galaxy clusters at low and mid-frequency with the SKA telescope}
   \subtitle{}

   \author{
   Valentina Vacca
   \inst{1}
\and
Federica Govoni
 \inst{1}
 \and
Matteo Murgia
\inst{1}
 \and
 Francesca Loi
\inst{1}
 \and
 Luigina Feretti
 \inst{2}
\and
Hui Li
 \inst{3}
\and
Elia Battistelli
 \inst{4}
\and
Torsten A. En{\ss}lin
 \inst{5}
\and
Paolo Marchegiani
 \inst{1}
             }

   \institute{
INAF-Osservatorio
Astronomico di Cagliari, Via della Scienza 5, I-09047 Selargius (CA), Italy\\
\and
INAF - Istituto di Radioastronomia, Via P. Gobetti 101, 40129 Bologna, Italy\\
\and
Theoretical Astrophysics, Los Alamos National Laboratory, Los Alamos, NM, USA\\
\and
Sapienza—University of Rome—Physics department, Piazzale Aldo Moro 5—I-00185, Rome, Italy\\
\and
Max Planck Institute for Astrophysics, Karl-Schwarzschildstr. 1, 85741 Garching, Germany }

   \date{Accepted September 25, 2004}

  \abstract
  {Understanding the magnetised Universe is a major challenge in modern astrophysics,
   and cosmic magnetism has been acknowledged as one of the science key drivers of the most ambitious radio instrument ever planned, the Square Kilometre Array telescope. }
   {With this work we aim to investigate the potential of the Square Kilometre Array telescope and its precursors and pathfinders in the study of magnetic fields in galaxy clusters and filaments through diffuse synchrotron radio emission. Galaxy clusters and filaments of the cosmic web are indeed unique laboratories to investigate turbulent fluid motions and large-scale magnetic fields in action and much of what is known about magnetic fields in galaxy clusters comes from sensitive radio observations.}
    {Based on cosmological magneto-hydro-dynamical (MHD) simulations, we predict radio properties (total intensity and polarisation) of a pair of galaxy clusters connected by a cosmic-web filament.} 
    {We use our theoretical expectations to explore the potential of polarimetric observations to study large-scale structure magnetic fields in the frequency ranges 50-350\,MHz and 950-1760\,MHz. We also present predictions for galaxy cluster polarimetric observations with the Square Kilometre Array precursors and pathfinders, such as the LOw frequency ARay,2.0 and the MeerKAT+ telescope.}
    {Our findings point out that polarisation observations are particularly powerful for the study of large-scale magnetic fields, since they are not significantly affected by confusion noise. The unprecedented sensitivity and spatial resolution of the intermediate frequency radio telescopes make them the favourite instruments for the study these sources through polarimetric data, potentially allowing us to understand if the energy density of relativistic electrons is in equipartition with the magnetic field or rather coupled with the thermal gas density. Our results show that low frequency instruments represent as well a precious tool to study diffuse synchrotron emission in total intensity and  polarisation.}

   \keywords{acceleration of particles -- polarisation -- magnetic fields -- galaxies: clusters: intracluster medium -- cosmology: simulations -- large-scale structure of Universe }

   \maketitle

%%%%%%%%%%%%%%%%%%%%%%%%%%%%%%%%%%%%%%%%%%%%%%%%%%

%%%%%%%%%%%%%%%%% BODY OF PAPER %%%%%%%%%%%%%%%%%%

\section{Introduction}

Turbulence and shock waves in merging galaxy clusters are likely to (re-)accelerate a pre-existing electron population embedded in a $\sim\mu$G intracluster magnetic field to ultra-relativistic energies ($\gamma\gtrsim 10^4$). These non-thermal components
can be observed in the radio domain in the form of faint ($\sim$0.1\,$\mu$Jy/arcsec$^2$ at 1.4 GHz) diffuse
steep-spectrum ($S_{\nu}\propto \nu^{-\alpha}$, with $\alpha\approx1-1.3$) synchrotron sources, called radio halos and relics,
respectively located at the centre and in the periphery of a fraction of merging galaxy clusters. Typically, radio halos show a correlation between the radio power and the cluster X-ray luminosity and mass, and a smooth and regular morphology similar to the distribution of the thermal plasma. The radio relics show also a correlation between their radio power and the cluster X-ray luminosity and are generally characterised by an elongated morphology (see, e.g., the reviews by \citealt{Feretti2012}, \citealt{vanWeeren2019} and references therein).

While radio relics show strong fractional polarisation \citep[e.g., ][]{Clarke2006,Bonafede2009a,vanWeeren2010,vanWeeren2012,Loi2017,Loi2020,Rajpurohit2020}, 
only three radio halos 
\citep{Govoni2005,Bonafede2009b,Girardi2016}
show polarised emission, and for one of them polarisation has been detected on larger scales than total intensity \citep[i.e., about $R_{\rm 500}$][]{Vacca2022}.  
Thanks to the high sensitivity, spatial and spectral resolution, future Square Kilometre Array (SKA) data should enable one to observe radio halo polarised emission in a larger number of galaxy clusters \citep{Govoni2013}. Stronger constraints on cluster magnetic fields are expected when polarised emission from both radio halos and background radio galaxies is detected \citep[e.g.,][]{Loi2019a}. 
Indeed, the detection of polarised emission of these extended diffuse sources combined with Faraday rotation of background and embedded radio galaxies permit
us to better investigate the intracluster magnetic field power spectrum. The study of the intracluster magnetic field through the combinations of these two observables has been performed to date only in two cases \citep[e.g., ][]{Govoni2006,Vacca2010}. %{Vogt2003,Govoni2006,Vacca2010,Kuchar2011}. 
A detailed study of the magnetic field properties through the polarimetric properties of radio halos or background radio galaxies, or both, has been conducted only in a limited sample of galaxy clusters. Current studies reveal that magnetic fields are characterised by central strengths of a few $\mu$G in merging galaxy clusters and up a few tens of $\mu$G in relaxed cool core clusters. Indications of a connection between the magnetic field strength and the thermal gas density at the centre of the cluster have been found \citep{Govoni2017}. 

Beyond galaxy clusters, the \textit{Planck} satellite \citep{Planck2013,Planck2016a} 
revealed the Sunyaev Zeldovich (SZ) effect associated with a bridge of plasma connecting two galaxy clusters, A399 and A401. By complementing \textit{Planck} observations with Atacama Cosmology Telescope (ACT) data, \cite{Hincks2022} derived that the size of the bridge is much more extended ($\sim 12$\,Mpc) than the projected separation in the plane of the sky ($\sim 3$\,Mpc).
In this system synchrotron emission has been detected through the observation of a double radio halo in the two clusters \citep{Murgia2010} and for the first time along the filament of matter connecting them \citep{Govoni2019} and as a consequence the presence of non-negligible magnetic fields has been inferred.
Another case of diffuse emission between two galaxy clusters has been later confirmed in the system A1758 \citep{Botteon2020}, as well as a bridge connecting a cluster and a group in the Shapley Supercluster \citep{Venturi2022}. From a statistical point of view, detections still waiting for confirmation have been presented in \cite{Vacca2018}. An analysis based on stacking techniques revealed an average magnetic field strength of 0.03\,$\mu$G$\,\leq\,$B$\,\leq\,$0.06\,$\mu$G along cosmic filaments from total intensity observations \citep{Vernstrom2021}, as well as polarised emission associated with accretion shocks from the cosmic web \citep{Vernstrom2023}.

Simulations of pairs of galaxy clusters and the filament connecting them are essential to explore the potential of the SKA telescope, its precursors and pathfinders polarimetric observations to study the details of magnetic fields in the large-scale-structure. 
In this paper we present synthetic images of a galaxy cluster pair similar to A399-A401, as expected with LOw Frequency ARray (LOFAR\,2.0), SKA-LOW, MeerKAT+, and SKA-MID.  
Properties of magnetic fields in low-density environments such as filaments mainly reflect the pristine magnetic field strength and structure in the Universe, and therefore their study is of paramount importance to shed light on the origin of cosmic magnetism.
 
The paper is organised as follows. In Sect.\,\ref{mhd} we
present MHD simulations (thermal density, magnetic field, and temperature) of a pair of colliding galaxy clusters, and we outline their similarity with the A399-A401 system. In Sect.\,\ref{syntheticimages} we present synthetic X-ray, SZ, and radio images of the simulated system and how these synthetic images compare to available literature data. In Sect.\,\ref{results},
we present expectations with the SKA telescope, its precursor and pathfinders respectively at mid- and low-frequencies in total intensity and polarisation under equipartition assumption between the magnetic field and the relativistic electrons %. 
and in the case that the relativistic electron distribution has an
energy density equal to 0.3 percent of the thermal one. 
Finally, in Sect.\,\ref{conclusions} we draw our conclusions. 

In the following, we use a $\Lambda$CDM cosmology with
$H_0$ = 72\,km/s/Mpc, $\Omega_0$ = 0.258 and $\Omega_{\Lambda}$ = 0.742. We presented this simulated galaxy cluster pair as observed at a redshift $z=0.073$, the same as the redshift of A399-A401. With this cosmology, at this distance, 1$^{\prime\prime}$ corresponds to 1.354\,kpc.

\section{MHD simulations}
\label{mhd}

The cosmological MHD simulation presented here is obtained with the ENZO code \citep{Collins2010} with adaptive mesh refinement (AMR) developed by the group of Hui Li at the Los Alamos National Laboratories, USA. 
The simulation runs from $z=30$ to $z=0$ and follows 
the evolution of the dark matter, baryonic matter, and magnetic fields. It uses an adiabatic equation of
state with a specific heat ratio $\Gamma=$5/3 and does not include
heating and cooling physics or chemical reactions. 
The magnetic fields are 
injected by active galactic nuclei (AGNs) at $z=2-3$ and then amplified and spread over Mpc-scales during the late stages of
the merger \citep{Xu2012}. A description of how the magnetic fields are injected is given in \cite{Xu2008,Xu2009}.
For the purposes of this work, a single snapshot of the MHD simulation is used. 
This snapshot of the simulation captures
a pair of galaxy clusters and the filament connecting them immediately
before the merging process begins. The output of the simulation
consists of a set of three-dimensional cubes of $\approx$(\,6.42\,Mpc)$^3$ with a cell size of $\approx$\,10.7\,kpc containing the intracluster medium (ICM) physical parameters:
temperature, thermal plasma density and magnetic field. 
Some characteristics of the simulated system are summarised in Table\,\ref{tab:A}. 
From the values in this table, we can compute the central magnetic to thermal gas energy density ratio, which is 0.2\,percent, 0.5\,percent and 0.4\,percent, respectively for C1, C2 and the filament. 
We note that \cite{Xu2010} showed that the distribution of the resulting intracluster magnetic field at low redshifts is not very sensitive to the
exact injection redshifts and to the injected magnetic energies. Moreover, the purpose of this work is not to shed light on the magnetic field strength or on the magneto-genesis process, but rather on the capabilities of the future radio instrumentation to detect in total intensity and polarisation systems with properties similar to those currently known.

\begin{figure*}
\sidecaption
\centering
\includegraphics[width=12 cm]{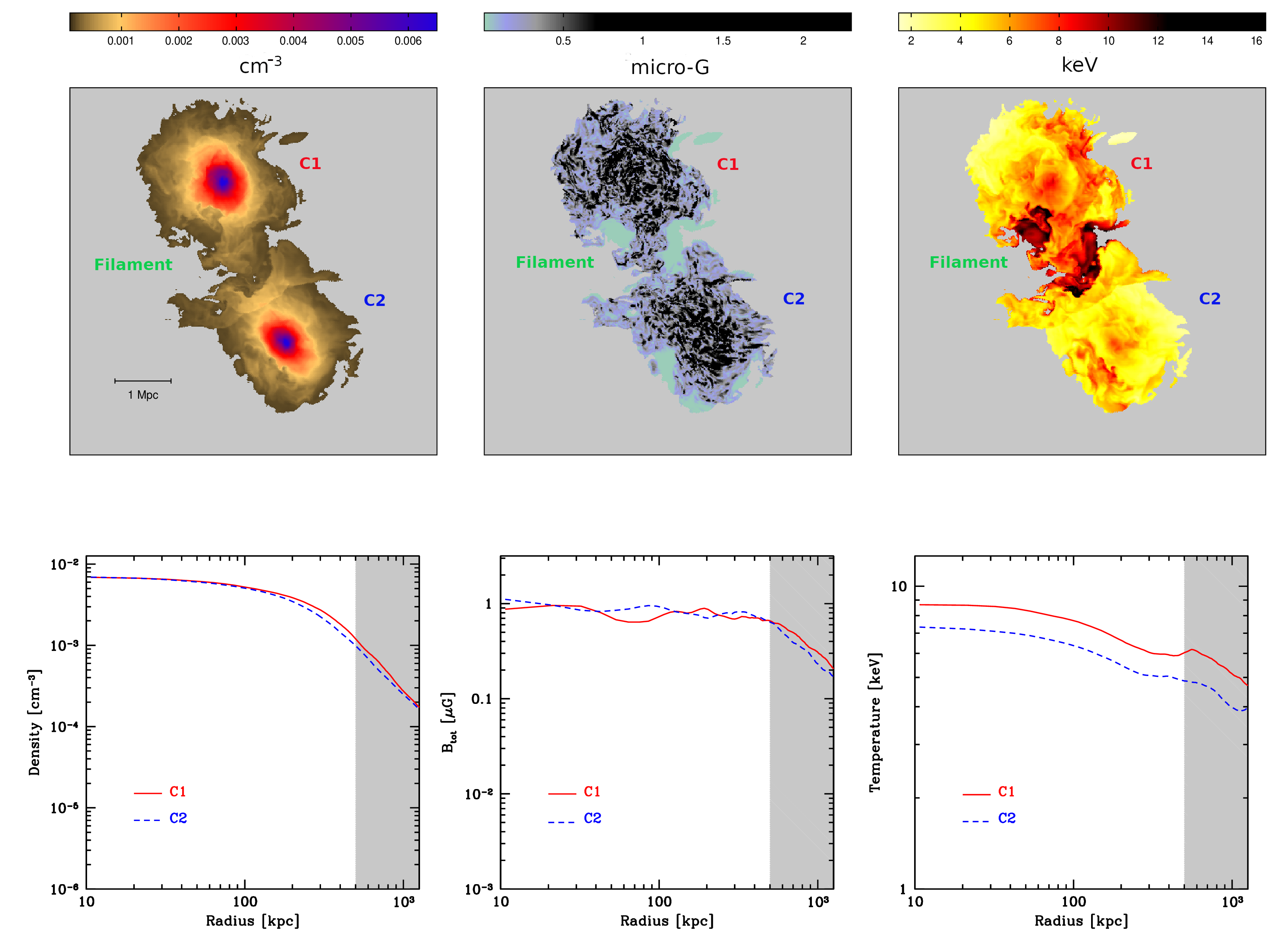}
\caption{
\emph{Top panels}: A central plane extracted from the cubes of the simulated thermal gas density (\emph{left}), total magnetic field (\emph{central}) and temperature (\emph{right}) of two merging galaxy clusters. Each image covers an area of 6.42\,Mpc $\times$ 6.42\,Mpc. 
\emph{Bottom panels}: Spherically averaged radial profiles of the simulated thermal gas density (\emph{left}), total magnetic field (\emph{central}) and temperature (\emph{right}) of the same system. These profiles are calculated in concentric spherical shells with a one-voxel width, 1\,voxel\,=(10.7\,kpc)$^3$, and starting from the cluster' centres. The grey shaded region indicates the radius above which the profiles must be taken with caution because the filament properties affect the profiles. 
}
\label{fig1}
\end{figure*}

In Fig.\,\ref{fig1}, we show in the top panels a central plane extracted from the cubes of the simulated thermal gas density, total magnetic field and temperature of two merging galaxy clusters. In the bottom panels we show the spherically averaged radial profiles computed on the simulated cubes of thermal gas density, total magnetic field and temperature of the same system. These profiles are calculated in concentric shells with a one-voxel width and starting from the clusters' centres, defined as the thermal gas density peak in three-dimensions. The three-dimensional distance between the two clusters is 3\,Mpc. 
We chose a mid-plane slice of that simulation passing close to the centres of both clusters, as can be inferred from Fig.\,\ref{fig1}. 
The two galaxy clusters are connected by a magnetised filament of matter with larger temperature than the surrounding, indicating that the galaxy clusters are approaching each other.
When spherically averaging thermal gas density, magnetic field and temperature of the system, the estimates of these quantities get contaminated by the filament volume.
Therefore, in the spherically averaged profiles we indicated with a grey shaded region the radius above which the profiles must be taken with caution because the filament properties affect the profiles. 
Moreover, we stress that the properties inferred from the spherically averaged radial profiles are not representative of the physical conditions in the filament, because they are derived by azimuthally averaging over a region of space including also the medium around the clusters.

\begin{table}
\caption{Properties of the system from the MHD simulation.}  
\label{electrons}      
\centering          
\begin{tabular}{ccccc}    
\hline     
     Cluster                       & $R_{500}$& $n_0$   &$B_0$& $T_0$ \\ %& $E_{\rm B_{\rm 0}}/E_{\rm th_{\rm 0}}$ \\ 
                         & kpc&    cm$^{-3}$ &$\mu$G&    keV  \\% & percent \\ 
\hline   
     C1                          &118     & $7.01\,\times\,10^{-3}$ & 0.79&  8.69   \\ %& 0.2  \\ 
     C2                          &107     & $7.10\,\times\,10^{-3}$ & 1.23 & 7.45  \\ %& 0.5  \\ 
     Filament                    &-     & $0.40\,\times\,10^{-3}$ & 0.30 & 8.35    \\ %& 0.4  \\ 
\hline  
\multicolumn{5}{l}{\scriptsize Col.\,1: Simulated cluster; Col.\,2: $R_{500}$. The $R_{500}$ radius is defined
as the distance within}\\ 
\multicolumn{5}{l}{\scriptsize which the cluster density is 500 times the critical density of the Universe; Col.\,3:}\\ 
\multicolumn{5}{l}{\scriptsize Central thermal electron density;  Col.\,4: Central magnetic field strength; Col.\,5:}\\ 
\multicolumn{5}{l}{\scriptsize Central temperature.}\\ 
\end{tabular}
\label{tab:A}
\end{table} 

\begin{figure*}
\sidecaption
\centering
\includegraphics[width=12 cm]{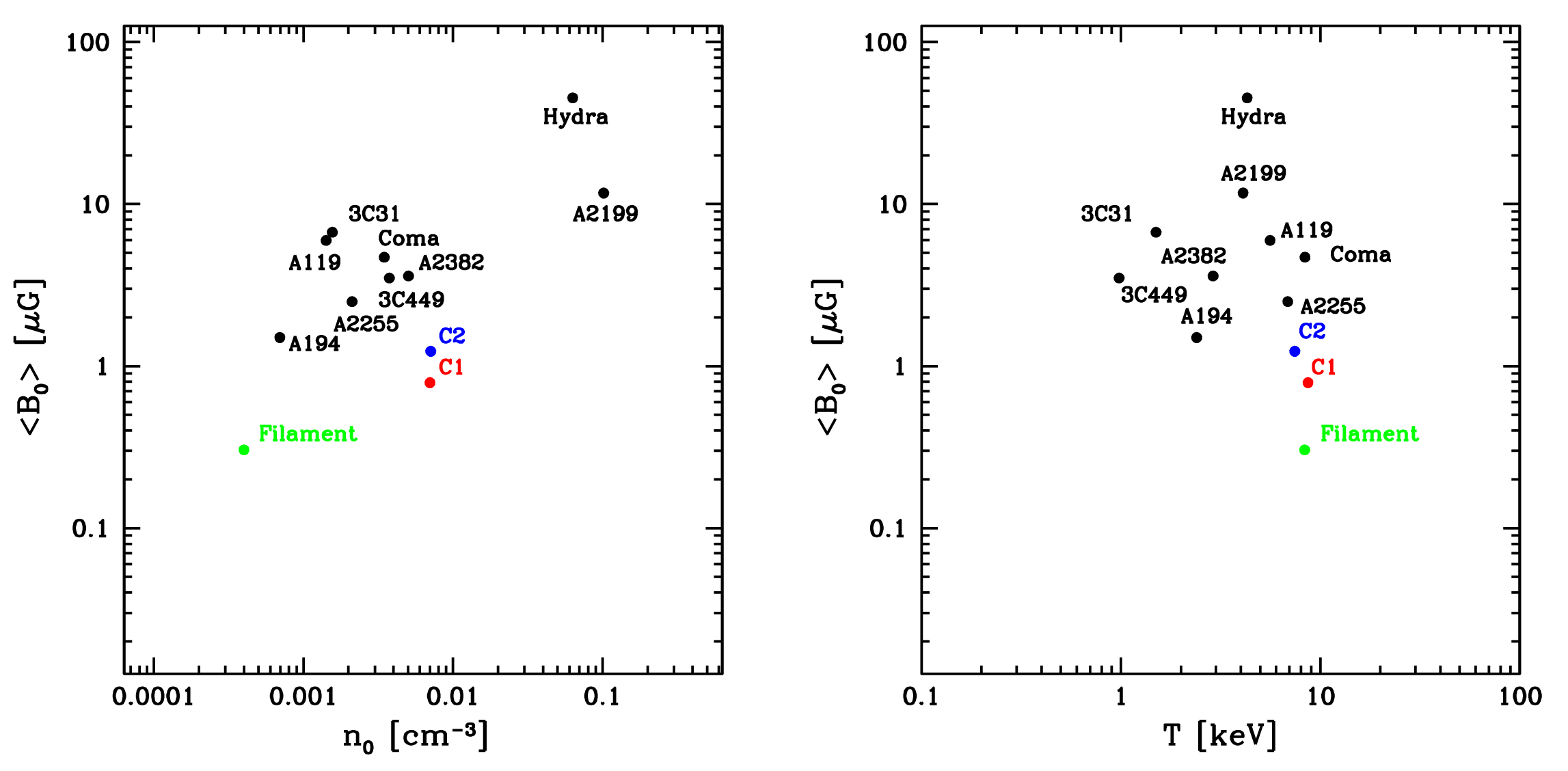}
\caption{
\emph{Left-hand panel}: Plot of the central magnetic field strength $\langle  B_{\rm 0}\rangle$ versus the central electron density $n_0$. \emph{Right-hand panel}: Plot of $\langle  B_{\rm 0}\rangle$ versus the mean cluster temperature T. For both plots, data points have been taken from \cite{Govoni2017}.}
\label{fig2}
\end{figure*}

It is interesting to compare the intracluster magnetic field of our simulations to that of real galaxy clusters, for which a detailed estimate is present in the literature.
In the left panel of Fig.\,\ref{fig2} we show a plot of the central magnetic field strength $\langle  B_{\rm 0}\rangle$\footnote{We note that $B_0$ refers to the magnetic field strength in the simulated cube at the voxel  corresponding to the peak in thermal gas density, while $\langle  B_{\rm 0}\rangle$ refers to the magnetic field at the centre of the cluster as derived from the comparison of observed and synthetic rotation measure images of galaxies in the cluster.} versus the central electron density $n_0$, while in the right panel of Fig.\,\ref{fig2} we show a plot of $\langle  B_{\rm 0}\rangle$ versus the mean cluster temperature T. 
These values have been taken from  \cite{Govoni2017} and have been converted to our cosmology. In general, fainter central
magnetic fields seem to be present in less dense galaxy clusters.
As noted in \cite{Govoni2017}, although the cluster sample is still rather small, 
we note that there is a hint of a positive trend between $\langle  B_{\rm 0}\rangle$ and $n_0$ measured among different clusters, while no correlation seems to be present between the central magnetic field and the mean cluster temperature. 
In the figures, we plot the values of the  magnetic field, thermal gas density and temperature at the centre of the simulated clusters C1 and C2, and in the middle point of the filament. 
However, the temperature of the simulated clusters is rather constant, as shown by the bottom right panel in Fig.\,\ref{fig1}. The simulated galaxy clusters are characterised by a weak central magnetic field, intermediate central thermal gas density and high temperature, in comparison with the clusters in the sample.

We selected this simulated system because its physical properties and configuration are similar to that observed for the pair of galaxy clusters A399–A401. Just to give a few numbers, the thermal gas densities at the centre of A399 and A401 are respectively $n_0=4.19\times10^{-3}$\,cm$^{-3}$ and $n_0=6.76\times10^{-3}$\,cm$^{-3}$, while  
the mean temperatures are $T=$7.23\,keV and 8.47\,keV \citep{Sakelliou2004}. 
In order to facilitate the comparison, we rotated the simulated galaxy cluster pair so that the system is in the plane of the sky and oriented as A399-A401.
As in our simulated system, X-ray, SZ, optical and radio data \citep{Sakelliou2004,Bourdin2008,Fujita2008,Murgia2010,Akamatsu2017,Bonjean2018,Govoni2019,Hincks2022} 
suggest that A399 and A401 are still in the initial phase of a merger, when the bulk of kinetic energy of the collision has not been dissipated yet.

\begin{figure*}
\centering
\includegraphics[width=14.5 cm]{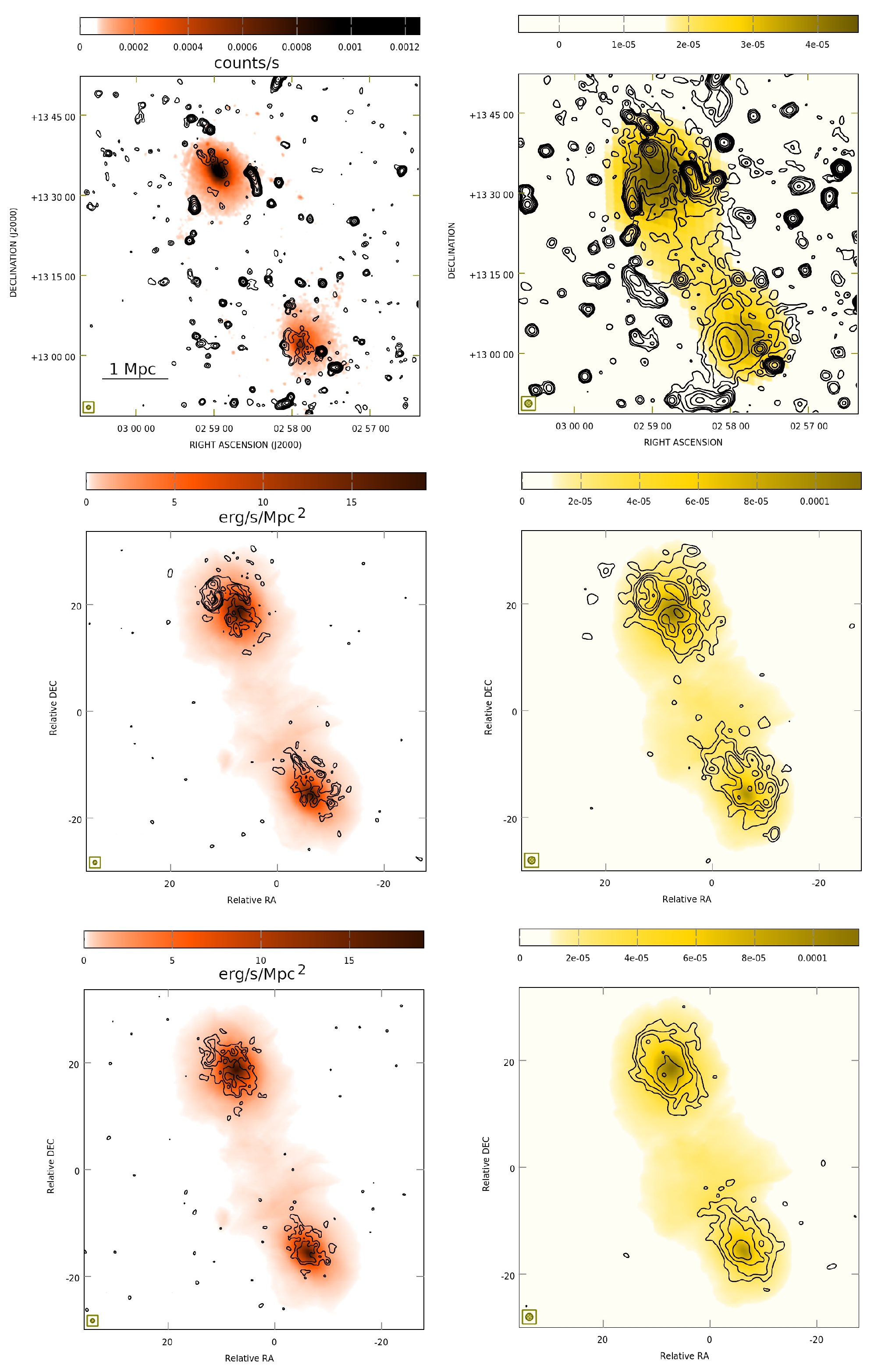}
\caption{
\emph{Top row}: Observations of the galaxy cluster pair A399-A401, with radio contours at about 1.4\,GHz over-imposed on the X-rays images in the band 0.2-12\,keV in red scale \citep{Murgia2010} on the left, and radio contours at about 150\,MHz \citep{Govoni2019} over-imposed on the Compton parameter images in yellow scale \citep{Hincks2022} on the right. \emph{Middle row}: The same as in the top row, but as produced from the simulations presented in this work, assuming equipartition between magnetic field and relativistic electrons at each point of the computational grid. 
 \emph{Bottom row}: The same as in the top row, but as produced from the simulations presented in this work, assuming a coupling between the relativistic and thermal electrons. 
In the left panels, contour levels start at 120\,$\mu$Jy/beam and increase by factors of 2, with a spatial resolution of 45$^{\prime\prime}$. In the right panels, contour levels start at 3\,mJy/beam and increase by factors of 2, with a spatial resolution of 80$^{\prime\prime}$. }
\label{fig3}
\end{figure*}

 \begin{figure*}
 \sidecaption
\centering
\includegraphics[width=12 cm]{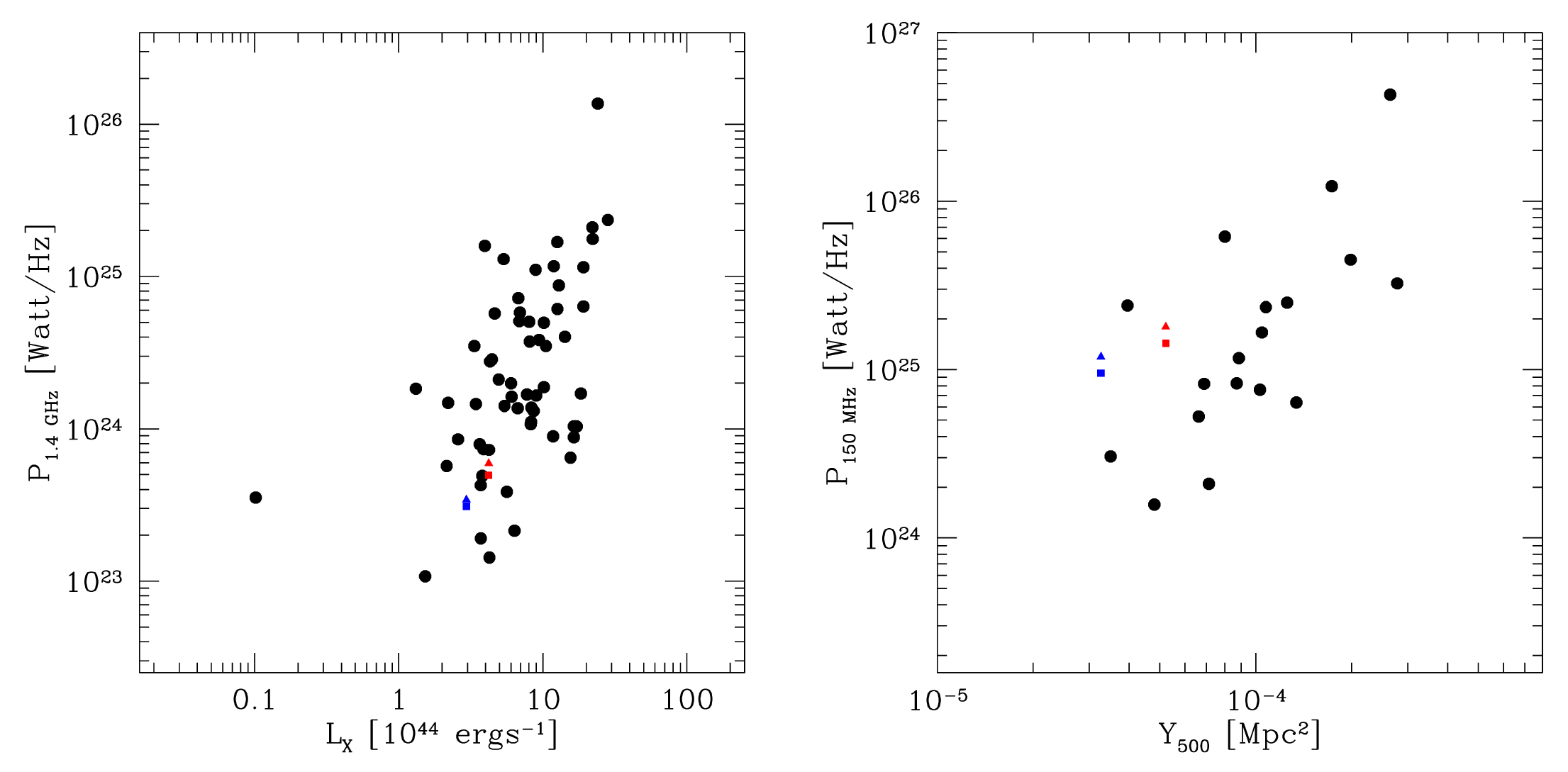}
\caption{\emph{Left-hand panel}: Radio power at 1.4\,GHz versus the X-ray luminosity in the 0.1-2.4\,keV energy range. The data are measurements for radio halos revealed through pointed interferometric observations at about 1.4\,GHz and are the same shown in \protect\cite{Loi2017}, plus the new halos from \protect\cite{Giovannini2020}. 
\emph{Right-hand panel}: Radio power at 150\,MHz versus the Compton parameter. The data are the literature measurements for radio halos revealed through pointed interferometric observations at about 150\,MHz reported by \cite{vanWeeren2021}.
In both plots, the values for the galaxy cluster C1 and C2 have been shown respectively in red and in blue, with triangles when derived assuming equipartition between magnetic field and relativistic electrons at each point of the computational grid and with squares when assuming a coupling between the relativistic and thermal electrons}.
\label{fig4}
\end{figure*}

\section{Synthetic X-ray, SZ and radio images}
\label{syntheticimages}
In this section we present synthetic X-ray, SZ, and radio images of the simulated system, obtained giving in input the thermal gas density, temperature and magnetic field cubes introduced in the previous section to the software package FARADAY \citep{Murgia2004}. The images have been produced in similar energy/frequency ranges as those available in the literature for A399-A401:

\begin{itemize}
\item we produced the X-ray images in the 0.1-2.4\,keV and 0.2-12\,keV energy range assuming a thermal bremsstrahlung model and by integrating the X-ray emissivity along the line-of-sight. The final output image is the X-ray brightness in the two energy ranges;

\item we produced the SZ effect by integrating the thermal gas density and temperature along the line of sight in order to produce the y-parameter image;

\item we produced synthetic radio images at the radio frequencies available in the literature (110-166\,MHz, 1.375-1.425\,GHz), and other frequency ranges expected with the SKA telescope, its precursor and pathfinders respectively at mid- and low-frequencies (i.e., LOFAR\,2.0, MeerKAT+, SKA-LOW and SKA-MID) by illuminating the magnetic field cube with a population of relativistic electrons. 
\end{itemize}
In the following we pay more attention to the generation of the radio images since the purpose of this work is to investigate the potential of the future radio instrumentation in detecting diffuse synchrotron radio emission from galaxy clusters and filaments. 
Following \cite{Murgia2004}, at each point on the computational grid,
we calculate the total intensity and the intrinsic linear
polarisation emissivity at each frequency, by convolving the emission spectrum of 
a single relativistic electron with the particle energy distribution 
of an isotropic population of relativistic electrons: 
$N(\gamma,\theta)=K_{\rm 0}\gamma^{-\delta}(\sin\theta)/2 $, with $\gamma_{\rm min}\,<\gamma\,<\gamma_{\rm max}$,
where $\gamma$ is the electron's Lorentz factor, and $\theta$ is the pitch 
angle between the electron's velocity and the local direction of the 
magnetic field\footnote{Please note that here $N(\gamma,\theta)$ stands for $\frac{{\mathrm d^2} N}{{\mathrm d \gamma}{\mathrm d \theta}}$}.
In Table\,\ref{tab2}, we indicate the parameters of the adopted distribution for
the relativistic particles. We consider two scenarios: one assuming equipartition between the magnetic field ($u_B$)
and the relativistic electron ($u_{el}$) energy density at every
location in the intracluster medium, and the other assuming a relativistic electron distribution with an energy density equal to 0.3\,percent of the thermal one ($u_{th}$). This factor
ensures a radio power at 1.4\,GHz consistent with that of radio halos in equipartition conditions and with the upper limit set from $\gamma$-ray observations, see \cite{Loi2019a} and \cite{Brunetti2017}. The slope of the electron energy distribution $\delta$ has been chosen in agreement with the spectral index $\alpha=1.6$ derived by \cite{Govoni2019}. The value for $\gamma_{\rm min}$ has been chosen in order to reproduce a radio halo luminosity at 1.4 GHz and 150 MHz in agreement with the correlation between the radio power and the cluster X-ray luminosity and the correlation between the radio power and 
the SZ Y-factor both in case of equipartition and in the case of relativistic electrons with an energy density equal to 0.3 percent of the thermal one, by keeping fixed the spectral index value to the value inferred by \cite{Govoni2019}.

The Stokes parameters $Q$ and $U$, the polarised 
intensity $P=(Q^2+U^2)^{1/2}$, and polarisation 
plane $\Psi=0.5\tan^{-1}(U/Q)$ images have been calculated by taking into account that the polarisation plane of the radio signal is subject to the Faraday rotation as it traverses the magnetised intracluster medium. 
Therefore, the integration of the polarised emissivity
along the line of sight has been performed 
as a vectorial sum in which the intrinsic polarisation angle of the radiation coming from the simulation cells located at a depth $L$ is rotated by an amount: 

\begin{equation}
 \Delta\Psi = \phi \times \left(\frac{c}{\nu}\right)^2, 
\label{rm1}
\end{equation}
where the Faraday depth $\phi$ is given by
\begin{equation}
\phi_{\rm~[rad/m^2]}=812\int_{0}^{L_{[kpc]}}n_{e~[cm^{-3}]}B_{\parallel~[\mu G]}dl.
\label{rm2}
\end{equation}
Here, $B_{\parallel}$ is the magnetic field along the line-of-sight. This effect leads to the so-called internal depolarization of the radio signal of the diffuse emission.

In Fig.\,\ref{fig3}, we compare the images of A399-A401 available in the literature (top panels) with the images obtained from the simulations both in case of equipartition of the relativistic electrons with the magnetic field (middle panels) and in case of a coupling of the relativistic and thermal electrons (bottom panels) at each point of the computational grid. In the simulation of the radio emission, we considered the diffuse emission only, because the radio galaxy emission as simulated by \cite{Loi2019a} is beyond the purposes of this analysis. To properly compare observations and simulations, we convolved the synthetic radio images for each Stokes parameter with the observing beam and added a Gaussian noise with rms consistent with the observations. In the left panels total intensity radio contours at about 1.4\,GHz are over-imposed on the X-rays images in the band 0.2-12\,keV in red scale, while in the right panels total intensity radio contours at about 150\,MHz are over-imposed on the Compton parameter images in yellow scale. While the LOFAR observations clearly show the presence of a radio ridge connecting A399 and A401, in our simulations only hints of diffuse emission between the two clusters can be identified. We stress that the simulated system is similar to A399-A401 in terms of thermal and non-thermal properties but, while the simulated system considered here is completely on the plane of the sky, A399-A401 is inclined along the line of sight \citep{Hincks2022}, which implies that the radio signal actually observed from the filament between
the two clusters may be larger as a result of a longer integration path along the line of sight. %meaning that the observed radio signal in between the two clusters is larger as a result of the integration along the line of sight.

In Fig.\,\ref{fig4}, we show, on the left, the radio power at 1.4\,GHz versus the X-ray luminosity in the 0.1-2.4\,keV energy range and, on the right, the radio power at 150\,MHz versus the Compton parameter. 
Black dots represent radio halos in the literature,
while the red and the blue colours the simulated radio halos respectively in the galaxy clusters C1 and C2 with triangles when assuming equipartition between magnetic field and relativistic electrons and squares when assuming a coupling between the relativistic and the thermal electron population. We computed the X-ray luminosity, and the radio flux density at 1.4\,GHz and at 150\,MHz for C1 and C2  within a box of radius 12.3\,arcmin (i.e., 1\,Mpc). This radius comprises the region where the radio brightness is above the 3$\sigma$ of the A399-A401 observations. Following \cite{vanWeeren2021}, we determined the $P_{\rm 150\,MHz}-Y_{\rm 500}$ radio halo scaling relation using the Compton Y parameter from the \textit{Planck} Sunyaev Zeldovich catalogue (PSZ2). For that, we converted the $Y_{\rm 5R500}$ values from \cite{Planck2016b} to $Y_{\rm 500}$
using $Y_{\rm 500} = 0.56 Y_{\rm 5R500}$ \citep{Arnaud2010}. We also converted the units from arcmin$^2$
to Mpc$^2$.

\begin{table*}[ht]
\caption{Sensitivities corresponding to different frequency ranges and channel widths considered in this work for the SKA telescopes and its precursors and pathfinders, see the text for more details.}      %\rotatebox{90}{
\centering          
\begin{tabular}{c c c c c c c c c c c}     
\hline\hline   
\multicolumn{1}{c}{Instrument} &\multicolumn{1}{c}{rms, 1\,hr, 10\,MHz}  &\multicolumn{3}{c}{$t_1=10$\,h} &\multicolumn{3}{c}{$t_2=100$\,h}\\ 
\hline
& &$\sigma_{\rm I, Q, U}$&$\sigma_{\rm P}$ &$\sigma_{\rm P, \phi}$ &$\sigma_{\rm I, Q, U}$&$\sigma_{\rm P}$ &$\sigma_{\rm P, \phi}$\\ 
&mJy/beam &$\mu$Jy/beam&mJy/beam &$\mu$Jy/beam &$\mu$Jy/beam&mJy/beam &$\mu$Jy/beam\\
\hline 

\multicolumn{8}{c}{}\\
\multicolumn{8}{c}{$\nu$=110-166\,MHz, d$\nu$=45\,kHz}\\
\hline
LOFAR\,2.0                      & 0.41         &54.9 &  5.9 & 77.6     & 17.3& 1.9& 24.5\\% &5.5&7.8&0.5\\

SKA-LOW                       & 0.047         &6.3  & 0.7 & 8.9   &2.0 & 0.2 &2.8 \\%&0.6&0.9&0.06\\
\hline 
\multicolumn{8}{c}{}\\
\multicolumn{8}{c}{$\nu$=110-166\,MHz, d$\nu$=12\,kHz}\\
\hline 

SKA-LOW                       & 0.047         & 6.3&    1.3 &8.9  & 2.0  & 0.4 &2.8\\%&0.6&0.9&0.1\\
\hline 
\multicolumn{8}{c}{}\\
\multicolumn{8}{c}{$\nu$=110-350\,MHz, d$\nu$=45\,kHz}\\
\hline 
SKA-LOW                       & 0.042         &2.7&     0.4& 3.8   & 0.9& 0.1& 1.2 \\%&0.3&0.4&0.02\\
\hline 
\multicolumn{8}{c}{}\\
\multicolumn{8}{c}{$\nu$=900-1670\,MHz, d$\nu$=1\,MHz}\\
\hline 

MeerKAT+                       &0.02          &0.8&   0.06&   1.1   & 0.2&  0.02& 0.3\\%& 0.08&0.1&0.005 \\
\hline
\multicolumn{8}{c}{}\\
\multicolumn{8}{c}{$\nu$=950-1760\,MHz, d$\nu$=1\,MHz}\\
\hline 

SKA-MID                       &0.013         &0.5  &   0.04&   0.6 & 0.1&  0.01& 0.2\\%&0.05&0.06&0.002 \\ 
&&&&&&&&&&\\
&&&&&&&&&&\\
\hline 
\hline
\multicolumn{1}{c}{Instrument} &\multicolumn{1}{c}{rms, 1\,hr, 10\,MHz}  &\multicolumn{6}{c}{$t=15$\,min} \\
\hline
& &\multicolumn{2}{c}{$\sigma_{\rm I, Q, U}$}&\multicolumn{2}{c}{$\sigma_{\rm P}$ }&\multicolumn{2}{c}{$\sigma_{\rm P, \phi}$}\\
&mJy/beam&\multicolumn{2}{c}{$\mu$Jy/beam} &\multicolumn{2}{c}{mJy/beam}&\multicolumn{2}{c}{$\mu$Jy/beam} \\
\hline
\multicolumn{8}{c}{}\\
\multicolumn{8}{c}{$\nu$=950-1760\,MHz, d$\nu$=1\,MHz}\\
\hline 
SKA-MID                       &0.013         &\multicolumn{2}{c}{2.9}&  \multicolumn{2}{c}{0.2}& \multicolumn{2}{c}{4.1}\\
%(2\,arcsec)                   &         &  &   &    & &  & \\%&0.05&0.06&0.002 \\ 

\hline
\hline

\end{tabular}
\label{sensitivity}
\end{table*}

\begin{table}
\caption{Confusion noise in the Stokes parameters I ($\sigma_{\rm c - I}$) and Q,U ($\sigma_{\rm c - QU}$) in the frequency ranges considered in this work \citep{Condon1974,Loi2019b}.}  \centering          
\begin{tabular}{c c c c}     
\hline\hline   
$\nu$&Resolution & $\sigma_{\rm c - I}$  &$\sigma_{\rm c - QU}$ \\
MHz  &arcsec     & $\mu$Jy/beam & $\mu$Jy/beam\\
\hline
110-166        & 20  &109       &  0.3    \\
               & 80  &2145      &  4.9    \\
110-350        & 80  &1425      &  3.3    \\
900-1670       & 20  &18.3      &  0.05   \\
               & 80  &360       &  0.8\\
950-1760       & 2   &0.1       &  0.0004 \\ 
               & 20  &17.5      &  0.05  \\ 
               & 80  &345       &  0.8    \\ 
               \hline
\hline
\end{tabular}
\label{confusion}
\end{table}

This analysis confirms that the amplitudes of the signals reproduced in the simulated images are consistent with the observations.
Moreover, the radio power (1.4\,GHz and 150\,MHz), Compton parameter and X-ray luminosity (0.1-2.4\,keV) of the simulated clusters C1 and C2 are in line with the correlations $P_{\rm 1.4\,GHz}-L_{\rm 0.1-2.4\,keV}$ and $P_{\rm 150\,MHz}-Y_{\rm 500}$ known from literature.

The most remarkable difference is that in case of equipartition we note a slightly prominent filamentary structure in the total intensity of the radio diffuse emission, as also suggested by \cite{Loi2019a}.

\section{Results and discussion}
\label{results}

In this section, we present our results. We predict the radio emission in total intensity and polarisation of a pair of galaxy clusters connected by a cosmic-web filament, similar to the system A399-A401 \citep{Govoni2019}, as expected with the SKA-LOW, SKA-MID, LOFAR\,2.0, and MeerKAT+ radio telescopes, in order to explore the potential of these instruments to study magnetisation of the large-scale-structure of the Universe. 
We produce our simulated images in different frequency ranges and with different frequency resolutions:
\begin{itemize}
    \item at low frequencies with LOFAR\,2.0 and SKA-LOW:
    \begin{itemize}
    \item[-] between 110 and 166\,MHz, with spectral resolution 12\,kHz and 45\,kHz;
    \item[-]  between 110 and 350\,MHz, with spectral resolution 45\,kHz;
    \end{itemize}
    \item at intermediate frequencies with MeerKAT+ and SKA-MID:
    \begin{itemize}
    \item[-]  between 900 and 1670\,MHz, with spectral resolution 1\,MHz;
    \item[-]  between 950 and 1760\,MHz, with spectral resolution 1\,MHz.
    \end{itemize}
\end{itemize}
In Table \ref{frequency} we present in a detailed way all the frequency coverages, bandwidths, spectral and spatial resolutions considered in this work. 

The SKA Magnetism Science Working Group identified as a top-priority a polarisation survey with SKA-MID \citep{Heald2020} with the following specifications: frequency range 950-1760\,MHz, sensitivity 4\,$\mu$Jy/beam, observing time 15\,min per pointing and spatial resolution 2$^{\prime\prime}$. Therefore, we adopt this observing setup here\footnote{We computed the sensitivity values in  Table\,\ref{sensitivity} according to \cite{Braun2019}, see the following.}.
At lower frequencies, discussion is still in progress to define the best observing setup and, for this reason, we consider here different configurations, i.e. different frequency ranges and different spectral resolutions. 
Full resolution images have been smoothed at 20$^{\prime\prime}$ and 80$^{\prime\prime}$, and for SKA-MID at 2$^{\prime\prime}$ as well. 
Noise has been added to 
the images as described in the following (see Sect.\,\ref{noise_estimate}).

After convolving with the instrumental full width half maximum (FWHM) and adding thermal and confusion noise, we performed rotation measure (RM) synthesis with the software FARADAY \citep{Murgia2004} on the resulting simulated Q and U cubes.  
To examine the results in output from the RM synthesis, we exploit the information contained in the polarisation image corresponding to the peak in the Faraday depth spectrum along each line of sight in the cube, after correcting for the polarisation bias according to \cite{George2012}
\begin{equation}
P^{\prime}=\sqrt{P^2-2.3\sigma_{\rm Q,U}^2},
\end{equation}
where $P$ and $P^{\prime}$ are respectively the polarised intensity before and after the bias correction, and $\sigma_{\rm Q,U}$ is the noise over the full band-with in Stokes Q and U.

In the images shown in the following we draw the contours in total intensity at 3$\sigma_{\rm I}$ while in polarisation at $5\sigma_{\rm P}$. In this work, we assume $\sigma_{\rm Q}=\sigma_{\rm U}$ that implies $5\sigma_{\rm P}\approx7\sigma_{\rm Q, U}$. A more severe cut in polarisation with respect to total intensity has been considered, since it translates into a false detection rate of 0.033\% \citep[see Table\,1 in ][]{George2012}.

\subsection{Noise estimate}
\label{noise_estimate}
When considering a data-set in the form of a frequency cube with $N_{\nu}$ channels with a thermal noise $\sigma_{\nu}$ per channel and per Stokes parameter, the thermal noise $\sigma_{\rm I, Q, U}$ per Stokes parameter averaged over the entire bandwidth is
\begin{equation}
\sigma_{\rm I, Q, U} =\frac{\sigma_{\nu}}{\sqrt{N_{\nu}}},
\end{equation}
and it is assumed to be the same for each Stokes parameter. 

If we apply RM synthesis, the noise over a channel of the Faraday depth cube depends on the noise in Q and U over the entire frequency band according to the following relation
\begin{equation}
\sigma_{\rm P, \phi}=\sqrt{2\sigma_{\rm Q,U}^2}=\sqrt{\frac{2\sigma_{\nu}^2}{N_{\nu}}}.
\end{equation}
For a Faraday depth cube consisting of $N_{\phi}$ slices, the noise {after summing over the entire width of the Faraday depth cube} is
\begin{equation}
\sigma_{\rm P}=\sigma_{\rm P, \phi}\sqrt{N_{\phi}}.
\label{sigma Faraday slice}
\end{equation}

In Table\,\ref{sensitivity} we report the expected sensitivity per single Stokes 
over all the frequency band ($\sigma_{\rm I,Q,U}$), the expected sensitivity of the full Faraday depth cube after summing all the channels ($\sigma_{\rm P}$) and the expected sensitivity in polarisation per single Faraday depth channel ($\sigma_{\rm P, \phi}$), 
 for all the frequency ranges, channel widths and integration times considered in this work and derived as described above. 
These values have been computed considering the reference values published in the \cite{LOFARWP2023} for LOFAR\,2.0, in the online documentation\footnote{\protect\url{https://www.meerkatplus.tel/mk-technical-details/}} for MeerKAT+, and in \cite{Braun2019} for the SKA telescope,  
and include the thermal noise only. 
In Table\,\ref{confusion} the confusion noise values for the same observing setups as in Table\,\ref{sensitivity}, and different spatial resolutions are given. These values have been computed according to \cite{Loi2019b}\footnote{$\sigma_{\rm c - QU}$ is an average of $\sigma_{\rm Q}$ and $\sigma_{\rm U}$.}, but see also \cite{Condon1974} for confusion noise estimates in total intensity only. 
When computing the sensitivity contours to be applied to the images, we included confusion noise $\sigma_{\rm c}$ by summing it in quadrature to 
the expected thermal sensitivity per single Stokes. E.g., the sensitivity over all the frequency band $\sigma_{\rm I, Q, U}^{\prime}$, is computed according to
\begin{equation}
    \sigma_{\rm I, Q, U}^{\prime}=\sqrt{\sigma_{\rm I, Q, U}^{2}+\sigma_{\rm c - I, QU}^2}.
\end{equation}

\begin{figure*}
 \sidecaption
\centering
\includegraphics[width=11 cm]{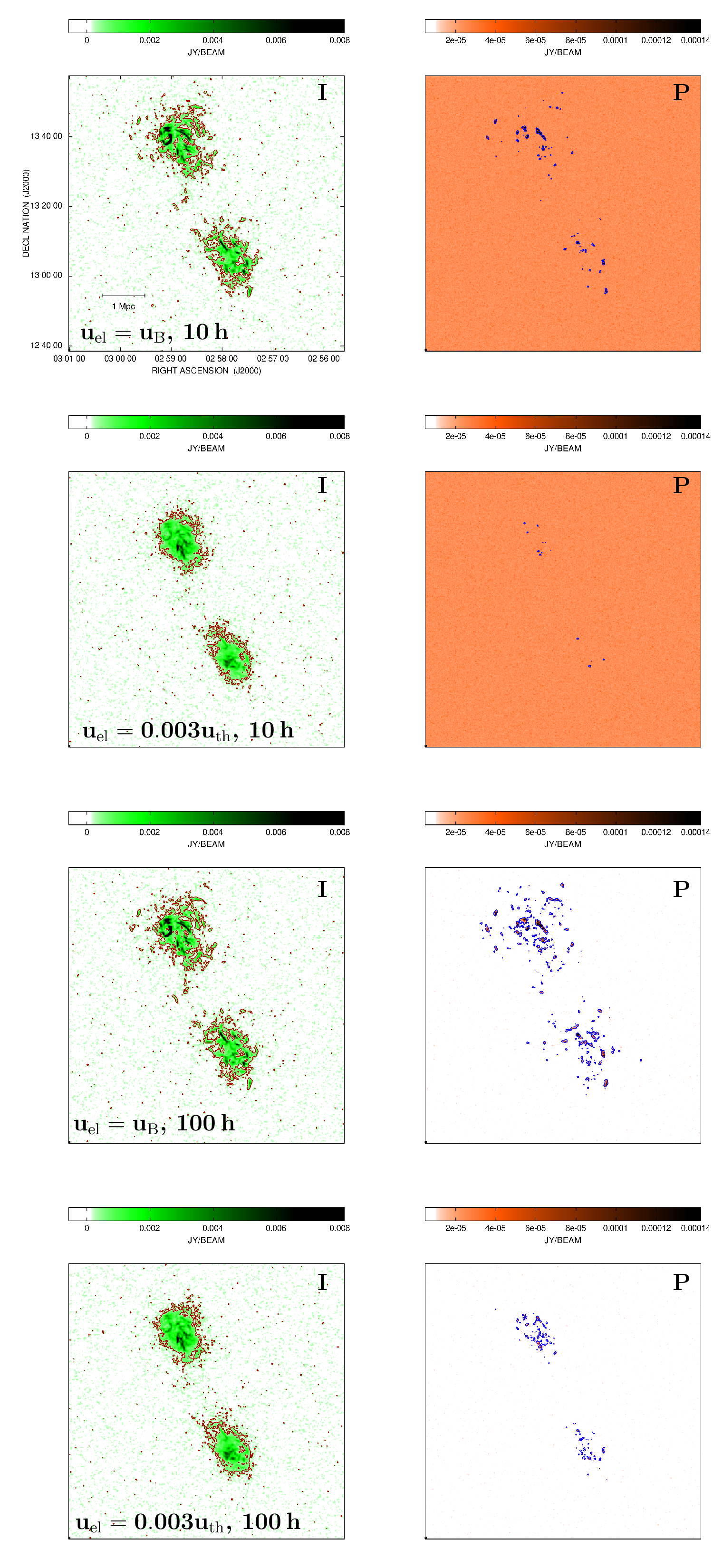}
\caption{Total intensity (\emph{left}) and polarised (\emph{right}) surface brightness images as expected from observations in the frequency range 110-166\,MHz and with a spectral resolution of 45\,kHz with the SKA-LOW telescope smoothed to a resolution of 20$^{\prime\prime}$. Red contours represent the 3$\sigma_{\rm I}$, while blue contours the  5$\sigma_{\rm P}$ sensitivity expected after 10\,h and 100\,h.
The FWHM is shown in the bottom left of each image. The simulated diffuse radio emission is obtained assuming a magnetic field and a thermal gas density distribution as simulated by \protect\cite{Xu2012} and a relativistic electron population in equipartition with the magnetic field (first and third row) and with an energy density equal to 0.3\,percent of the thermal one (second and fourth row), see the text for more details.}
\label{fig5}
\end{figure*}

%\begin{figure*}[ht!]
%\sidecaption
%\centering
%\includegraphics[width=11 cm]{Fig6.png}
%\caption{As Fig.\,\ref{fig5} but for SKA-LOW observations smoothed to a resolution of 80$^{\prime\prime}$.} 
%\label{fig6}
%\end{figure*}

%\begin{figure*}[ht!]
% \sidecaption
%\centering
%\includegraphics[width=11 cm]{Fig7.png}
%\caption{As Fig.\,\ref{fig5} but for LOFAR\,2.0 observations smoothed to a resolution of 20$^{\prime\prime}$.} 
%\label{fig7}
%\end{figure*}

%\begin{figure*}[ht!]
% \sidecaption
%\centering
%\includegraphics[width=11 cm]{Fig8.png}
%\caption{As Fig.\,\ref{fig5} but for LOFAR\,2.0 observations smoothed to a resolution of 80$^{\prime\prime}$. }
%\label{fig8}
%\end{figure*}

\subsection{Results at low frequencies}
In Fig.\,\ref{fig5} and Fig.\,6\footnote{Figures 6 to 13, B.1 and B.2 are available on Zenodo, please refer to the \emph{Data availability} section in this paper.}, we show the total intensity and polarization images obtained at low frequencies with SKA-LOW in the frequency range 110--166\,MHz with a spectral resolution of 45\,kHz at 20$^{\prime\prime}$ and 80$^{\prime\prime}$. Results for 10 and 100 observing hours, for the equipartition scenario and for a coupling between the relativistic and the thermal electrons, are shown. Please, note that we use the same colour-bar range respectively for all the total intensity and all the polarization images for easier comparison. In case of relativistic electrons coupled with thermal electrons, the radio halo shows a smoother morphology, with less filamentary substructures and a fainter polarized emission with respect to the equipartition scenario.
According to our simulations, SKA-LOW\footnote{We considered the sensitivity at 110\,MHz \citep{Braun2019}.} synthetic images at 80$^{\prime\prime}$ in total intensity allow us to image only the emission at the centre of the two clusters, while at 20$^{\prime\prime}$ marginal hints of emission along the filament connecting the two clusters can be identified as well. 
By increasing the observing time, we note that deeper observations do not improve the sensitivity and, consequently, do not reveal fainter structures in total intensity. Low frequency total intensity observations are indeed limited by confusion noise after a few seconds or minutes, according to the selected spatial resolution. 

Because of depolarisation, polarisation images at these frequencies allow us to detect only the brightest peaks of emission at the centre of the two clusters, at the periphery, and along the filament connecting them. Due to the lower density of polarised sources with respect to total intensity, confusion noise has a lower impact on polarimetric observations. Deeper  observations are therefore precious to reveal filamentary polarised structures both associated with the two clusters and with the environment in between them, even where a total intensity counterpart has been not detected.
For both spatial resolutions, a better detection of the diffuse emission is possible when considering equipartition between relativistic electrons and magnetic field.

In Appendix\,\ref{appendixB}, we explore the possibility to use low-frequency SKA-LOW data at high spectral resolution or over a larger bandwidth. The setup with higher spectral resolution (Fig.\,B.1), i.e. 12\,kHz, does not show significant differences with respect to Fig.\,\ref{fig5}, indicating that no bandwidth depolarisation is taking place or, if present, its contribution is negligible.
Expectations for observations over a larger frequency band, i.e. 110--350\,MHz  (Fig.\,B.2), 
look only slightly better. By comparing SKA-LOW images over 110-166\,MHz and over 110-350\,MHz it is evident that, at these frequencies and spatial resolution, confusion noise is the dominating noise source also in polarisation. 

As a comparison, we show in Fig.\,7 and in Fig.\,8 the total intensity and polarisation images obtained at low frequencies with LOFAR\,2.0 in the frequency range 110--166\,MHz with a spectral resolution of 45\,kHz at 20$^{\prime\prime}$ and 80$^{\prime\prime}$. Our simulations indicate that this instrument is capable of recovering comparable total intensity emission images as SKA-LOW, likely due to the fact that the noise is dominated by the confusion of background sources. However, in polarisation, the performance of LOFAR\,2.0 are worse than those of SKA-LOW. In the equipartition scenario, only a few bright peaks of emission can be detected. These peaks are almost buried by the noise when sensitivities corresponding to 10\,h of observations are considered, requiring an observing time of at least 100\,h in order to be detected. By comparing LOFAR\,2.0 and SKA-LOW images, the recovery of the signal after 100\,h with LOFAR\,2.0 is less effective than after only 10\,h with SKA-LOW.

%\begin{figure*}[ht!]
% \sidecaption
%\centering
%\includegraphics[width=11 cm]{Fig9.png}
%\caption{As Fig.\,\ref{fig5} but for SKA-MID observations in the frequency range 950--1760\,MHz, with a spectral resolution of 1\,MHz, and smoothed to a resolution of 20$^{\prime\prime}$.}
%\label{fig9}
%\end{figure*}

%\begin{figure*}[ht!]
% \sidecaption
%\centering
%\includegraphics[width=11 cm]{Fig10.png}
%\caption{As Fig.\,\ref{fig5} but for SKA-MID observations in the frequency range 950--1760\,MHz, with a spectral resolution of 1\,MHz, and smoothed to a resolution of 80$^{\prime\prime}$.}
%\label{fig10}
%\end{figure*}

%\begin{figure*}[ht!]
% \sidecaption
%\centering
%\includegraphics[width=11 cm]{Fig11.png}
%\caption{As Fig.\,\ref{fig5} but for SKA-MID observations in the frequency range 950--1760\,MHz, with a spectral resolution of 1\,MHz, and smoothed to a resolution of 2$^{\prime\prime}$.}
%\label{fig11}
%\end{figure*}

%\begin{figure*}[ht!]
% \sidecaption
%\centering
%\includegraphics[width=11 cm]{Fig12.png}
%\caption{As Fig.\,\ref{fig5} but for MeerKAT+ observations in the frequency range 900--1670\,MHz, with a spectral resolution of 1\,MHz, and smoothed to a resolution of 20$^{\prime\prime}$.}
%\label{fig12}
%\end{figure*}

%\begin{figure*}[ht!]
%\sidecaption
%\centering
%\includegraphics[width=11 cm]{Fig13.png}
%\caption{As Fig.\,\ref{fig5} but for MeerKAT+ observations in the frequency range 900--1670\,MHz, with a spectral resolution of 1\,MHz, and smoothed to a resolution of 80$^{\prime\prime}$.}
%\label{fig13}
%\end{figure*}

\subsection{Results at intermediate frequencies}

In Fig.\,9 and Fig.\,10, we show the results obtained with SKA-MID in the frequency range 950--1760\,MHz (band 2) with a spectral resolution of 1\,MHz at a spatial resolution respectively of 20$^{\prime\prime}$ and 80$^{\prime\prime}$.
The detection of total intensity emission is considerably hindered by the confusion noise. Indeed, only the brightest patches at the centre of the two clusters can be revealed. At a low spatial resolution of 80$^{\prime\prime}$, confusion noise dominates already after 10\,h both in total intensity and in polarisation. At 20$^{\prime\prime}$, polarimetric data prove to be much more powerful than total intensity ones. By increasing the observing time, indeed, they enable a good mapping of the polarised emission also toward the cluster outskirts and in between the two clusters, in regions where we do not detect the total intensity counterpart and therefore particularly precious. As these images clearly show, the most important result of the paper is that the sensitivity reached in polarisation is higher than in total intensity due to a lower confusion noise. This allows us to detect the polarised emission permeating the clusters and the filament between them. Both Fig.\,9 and Fig.\,10 indicate that a better imaging of the polarised emission is possible in the equipartition scenario.

In order to explore the potential of SKA-MID polarimetric survey planned by the SKA Magnetism Science Working group \citep[][see also text above]{Heald2020} to study magnetic fields in galaxy clusters and beyond through diffuse synchrotron emission total intensity and polarisation observations, we produced synthetic images corresponding to an observing time of 15\,min and to a spatial resolution of 2$^{\prime\prime}$, see Fig.\,11. With these specifications, we are able to detect total intensity patches similar to those identified at lower spatial resolution, since all these images are dominated by confusion noise. Concerning the polarised emission, we can detect only the brightest patches and only in the equipartition scenario. When relativistic and thermal particles are coupled, indeed, the expected signal is below the noise level.

In Fig.\,12 and Fig.\,13, we show the expectations with MeerKAT+ observations in the frequency range 900--1670\,MHz with a spectral resolution of 1\,MHz  at a spatial resolution respectively of 20$^{\prime\prime}$ and 80$^{\prime\prime}$. 
We note that the frequency range is similar to that considered for SKA-MID.
For this reason, for comparable observing times, the results are similar to those obtained for SKA-MID, but with better performances for the last, thanks to the higher sensitivity of this instrument. Please, note that in Fig.\,12 and Fig.\,13 we use the same colour-bar ranges as in Fig.\,9 and Fig.\,10, respectively, for easier comparison.

\subsubsection*{Overall considerations}

According to our results, the best observing instrument in order to study diffuse synchrotron emission is SKA-MID in the frequency range 950--1760\,MHz with a spectral resolution of 1\,MHz. While total intensity observations are heavily limited by the confusion of background radio sources within the observing beam, polarisation observations are only marginally affected by it. In particular, total intensity observations enable only the imaging of the central regions of the galaxy clusters. In order to map the emission over all the cluster volume and along the filament connecting the two clusters, polarimetric observations are crucial. Observing times of about 100\,h are necessary when spatial resolutions of 20$^{\prime\prime}$ are considered, while shorter observing times of about 10\,h are sufficient at 80$^{\prime\prime}$.
The combination of sensitivity and spatial resolution of intermediate frequency polarimetric observations allow us to map the detailed morphology of the diffuse emission and possibly track its filamentary structure. Because of this, according to our results, future observations can allow us to distinguish between the equipartition scenario and the scenario based on a coupling of thermal and relativistic electrons. In the first scenario, indeed, the diffuse radio emission is expected to be more filamentary, with a less smooth morphology, and with a higher radio brightness with respect to the second scenario and a consequent better imaging of the peripheral regions of the system. This is crucial in order to characterise the magnetic field and the relativistic electrons energy spectrum and their spatial distribution in a detailed way.

Our predictions for SKA-MID observations offer important food for thought also about what we should expect from the polarisation survey planned for SKA-MID in the frequency range 950 to 1760\,MHz, with a FWHM of 2$^{\prime\prime}$ and an observing time of 15\,min \citep{Heald2020}. According to our results, this survey will be not effective for the observation of the polarised diffuse emission associated with galaxy clusters and filaments of the cosmic web, which rather requires deeper pointed observations at lower spatial resolution.

Although the better sensitivity of the SKA-MID makes this the favourite instrument, MeerKAT+ performance appear to be good enough to already conduct this kind of studies, requiring observing times of about 100\,h for mapping the diffuse emission over a significant fraction of the cluster volume. 

The results presented in this work reveal that low frequency instruments represent as well a precious tool to study diffuse synchrotron emission, not only in total intensity but also in polarisation.
Even if low frequency observations are more deeply affected by the Faraday rotation, SKA-LOW observations are expected to detect the brightest patches of polarised emission associated with the centre of galaxy clusters as well as hints of it toward their periphery and between them, with better results for low spatial resolutions. Larger bandwidths and longer observing times perform only slightly better because at these frequencies and spatial resolutions confusion noise starts to dominate.
The nominal SKA-LOW frequency range starts from 50\,MHz, however in this work we explore only the frequency range down to 110\,MHz with a spectral resolution of both 45\,kHz, a factor two better than that currently used with published LOFAR polarisation works \citep[see, e.g.,][]{Herrera2021} and comparable with that one adopted for the analysis of new deep LOFAR observations (Vacca et al. in prep.). % and 12\,kHz. 
A spectral resolution of 12\,kHz would allow us to recover at least 90\,percent of the polarised emission of 100\,rad/m$^2$-sources \citep[see fig.\,10.6 in][]{Heald2018}. To exploit all the frequency range accessible with the SKA-LOW, ensuring an almost complete recovery of the polarized intensity would require a frequency resolution of at least 6\,kHz. The combination of large-bandwidth and high spectral resolution of this observing setup is computationally very expensive. Moreover, our results indicate that changing the spectral resolution from 45\,kHz to 12\,kHz does not allow us a significant better recovery of the signal. This indicates that the in-band depolarisation is negligible and that the contribution to the polarised brightness from high rotation measure sources is not relevant here. 

\setcounter{figure}{13}
\begin{figure}
\centering
\includegraphics[width=9 cm]{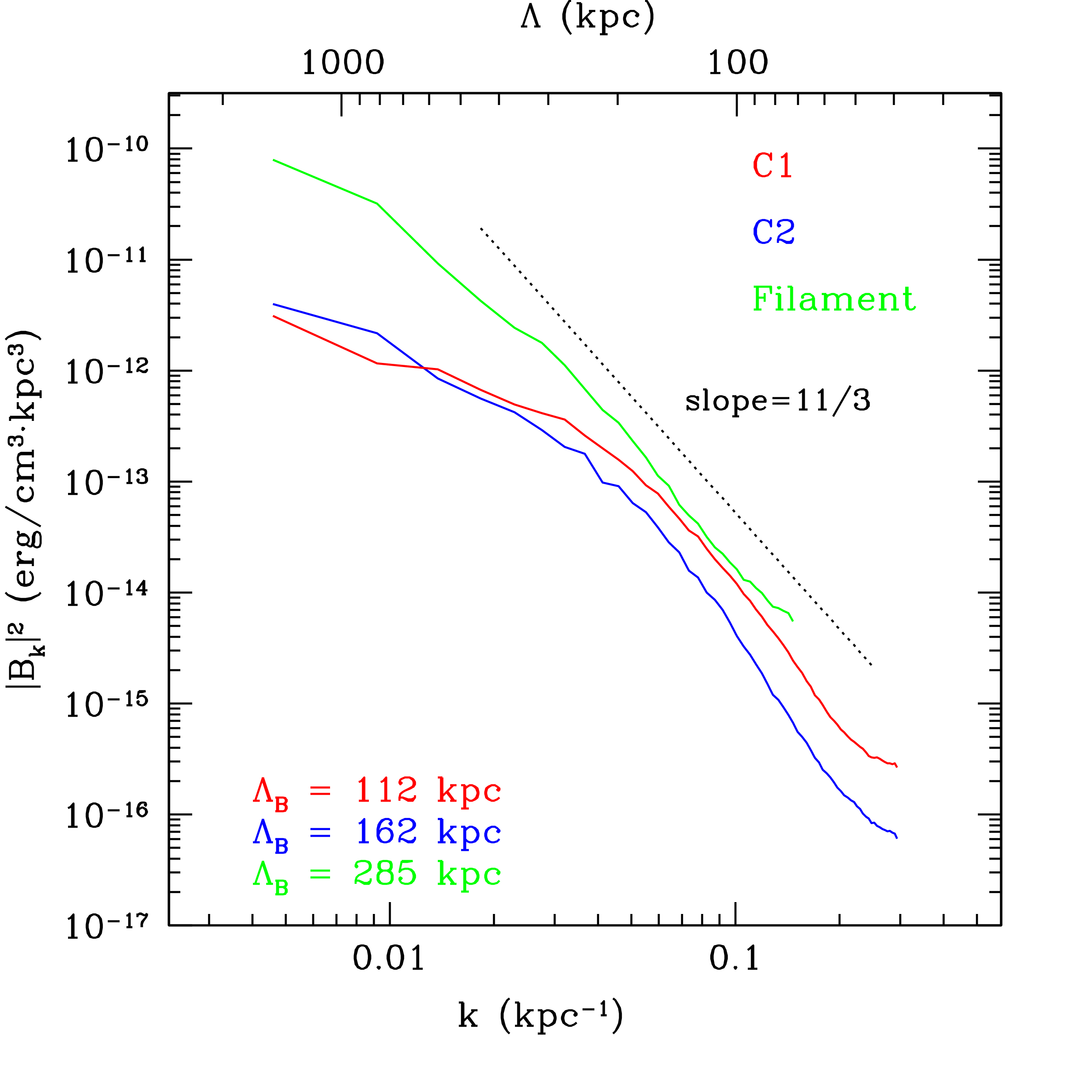}
\caption{Amplitude of the magnetic field power spectrum versus the wave number and the scale, for two galaxy clusters C1 (red) and C2 (blue) and for the filament (green). As a reference, the slope of a Kolmogorov magnetic field power spectrum is shown. 
}
\label{fig14}
\end{figure}

 Generally speaking, the polarised emission is fainter than the total intensity one and therefore more elusive, since the intrinsic polarisation degree is $\approx$\,75 -- 80\,percent of the total intensity signal, for typical values of the particle spectral index. However, since the total intensity is more affected by the confusion noise than the polarised emission, our simulations indicate that deep polarimetric observations can permit us to detect a signal without a detectable total intensity counterpart. 
Intermediate frequencies appear to be affected by this phenomenon as well, as also observed for the first time in a galaxy cluster system, see \cite{Vacca2022}. Even if, at these frequencies, the expected emission of the sources considered in this work is fainter than at lower frequencies due to the steep spectral index $\alpha\gtrsim 1$, simultaneously, the Faraday rotation has a lower impact, facilitating the detection of polarised emission. 

Additionally, the study of polarised emission associated with these systems can be sometime hindered by the depolarisation of the signal within the observing beam. Our results show that polarised emission can be detected for these systems even at very low spatial resolution (80$^{\prime\prime}$, i.e. 110\,kpc at the distance of the system), suggesting that likely the magnetic field is fluctuating over large scales in our simulations. In order to shed light on this, we compute the magnetic field power spectrum in the two clusters and along the filament, as shown in Fig.\,\ref{fig14}. We derived for each cluster the auto-correlation length $\Lambda_{\rm B}$ by applying eq.\,10 in \cite{Vacca2012}
\begin{equation}
\Lambda_{\rm B}=\frac{3\pi}{2}\frac{\int_0^{\infty}|B_{\rm k}|^2k\mathrm{d}k}{\int_0^{\infty}|B_{\rm k}|^2k^2\mathrm{d}k},
\end{equation}
see also references therein. We find respectively for the clusters C1 and C2, $\Lambda_{\rm B}=112$\,kpc and $\Lambda_{\rm B}=162$\,kpc, and for the filament $\Lambda_{\rm B}=285$\,kpc, larger than the lowest spatial resolution considered in this work. These findings indicate that the beam depolarisation is not affecting our results significantly since the magnetic field is ordered on scales larger than the beam.

\section{Conclusions}
\label{conclusions}

In this work, we use cosmological
magneto-hydro-dynamical simulations 
to predict the expected surface brightness distribution
of radio halos both in total intensity and in polarization
with next generation facilities, as SKA-LOW, SKA-MID, LOFAR\,2.0 and MeerKAT+.
Under a reasonable shape
for the relativistic electron energy spectrum, 
we produced polarimetric synthetic radio images at low (SKA-LOW and LOFAR\,2.0) and intermediate (SKA-MID and MeerKAT+) frequencies of a pair of approaching galaxy clusters similar to the system A399-A401 and with properties in agreement with systems hosting the currently known radio halos, at radio, X-rays and mm/sub-mm wavelengths. Our simulations indicate that while total intensity emission is usually dominated by confusion noise, polarized emission is less affected by it. For this reason, polarized emission can be detected also at locations where the total intensity is buried below the noise and therefore represents a powerful instrument to study the non-thermal components of galaxy clusters and of the cosmic web. %

Our results show that in order to better reconstruct the morphology of the diffuse emission in polarisation over a significant fraction of the volume of the system and to put constraints on the spatial distribution of the non-thermal components, deep observations at few tens of arcsec of resolution at intermediate frequencies appear to be the favourite option. Although the better sensitivity of the SKA-MID makes this the favourite instrument, MeerKAT+ performances appear to be good enough to already conduct this kind of study, but require at least one hundred hours of observation. 

Low frequency instruments represent as well a precious tool to study the brightest patches of diffuse synchrotron emission, in total intensity and polarisation in the centre of galaxy clusters and between them. 
Deep and low-spatial resolution observations with SKA-LOW proved to be more effective, provided that the auto-correlation length of the magnetic field is larger than the observing beam. On the other side, the capabilities of LOFAR\,2.0 do not appear to be suitable for this kind of studies.

Our findings are similar if we consider an equipartition scenario as well as in the case of a relativistic electron distribution with an energy density equal to 0.3\,percent of the thermal one. However, in the last case, the diffuse radio emission shows a less filamentary and smoother morphology, overall with a fainter radio brightness and therefore more elusive. 
Due to the combination of unprecedented resolution and sensitivity, future radio observations will allow us to characterise the properties of the diffuse synchrotron radio emission, potentially discriminating among scenarios assuming equipartition of magnetic fields and relativistic electrons and scenarios based on a coupling between thermal and non-thermal electrons. The comparison of the expectations presented here with observations that will be performed with the new generation of radio instruments will be crucial to shed light not only on the magnetic field properties but also on the distribution and energy content of the relativistic particles responsible for the diffuse synchrotron emission.

\section{Data availability}
Figures 6 to 13, B.1 and B.2 can be found at the following link\footnote{\protect\url{https://zenodo.org/records/13912170}}.

\begin{acknowledgements}
We thank the anonymous referee for the precious suggestions that helped to improve the quality of the paper.
V. V. acknowledges support from the Prize for Young Researchers "Gianni Tofani" second edition, promoted by INAF-Osservatorio Astrofisico di Arcetri (DD n. 84/2023).
FL and PM acknowledge financial support from the Italian Ministry of University and Research – Project Proposal CIR01-00010.
\end{acknowledgements}



%\bibliographystyle{mnras}


%%%%%%%%%%%%%%%%%%%%%%%%%%%%%%%%%%%%%%%%%%%%%%%%%%

%%%%%%%%%%%%%%%%% APPENDICES %%%%%%%%%%%%%%%%%%%%%

\begin{appendix}
\onecolumn

\section{Additional tables} 
\label{appendixA}

\begin{table*}[htp] 
\caption{Adopted parameters for the relativistic electron distribution.}  
\label{tab2}      
\centering          
\begin{tabular}{c l l}    
\hline       
    Parameter     & Value                    & Description          \\ 
\hline   
  $\gamma_{\rm min}$   & 300            & Minimum relativistic electron Lorentz factor \\
  $\gamma_{\rm max}$   & 1.5$\times$10$^{4}$ & Maximum relativistic electron Lorentz factor \\
  $\delta$        &    4.2         & Power-law index of the energy spectrum of the  \\
                  &                & relativistic electrons \\
  $K_0$           & Adjusted to guarantee $u_{\rm el}=u_{\rm B}$ and $u_{\rm el}=0.003u_{\rm th}$ & Electron spectrum normalization \\ 
                  &       at each point of  the computational grid       &  \\ 
\hline                                 
\end{tabular}
\end{table*}

\begin{table*}[htp]
\caption{Instrumental setup used in the simulations: frequency coverages, bandwidths, spectral and spatial resolutions 
considered in this work.}       \centering          
\begin{tabular}{c c c c c c}     
\hline\hline       
Instrument  & Frequency & Bandwidth & Channel width & Resolution & Figure \\
            & MHz     & MHz    &  MHz        & arcsec  & \\
\hline 
SKA-LOW         & 110$-$166   &  56 &  0.045        & 20  & Fig.\,\ref{fig5}\\
SKA-LOW         & 110$-$166   &  56 &  0.045        & 80  & Fig.\,6\\ 
SKA-LOW         & 110$-$166   &  56 &  0.012        & 80  & Fig.\,B.1\\ 
SKA-LOW         & 110$-$350   &  240 & 0.045        & 80  & Fig.\,B.2\\
LOFAR\,2.0      & 110$-$166   &  56 &  0.045        & 20  & Fig.\,7\\ 
LOFAR\,2.0      & 110$-$166   &  56 &  0.045        & 80  & Fig.\,8\\ 
SKA-MID         & 950$-$1760  & 810 &  1            & 20  & Fig.\,9\\
SKA-MID         & 950$-$1760  & 810 &  1            & 80  & Fig.\,10\\ 
SKA-MID         & 950$-$1760  & 810 &  1            & 2  & Fig.\,11\\ 
MeerKAT+        & 900$-$1670  & 770 &  1            & 20  & Fig.\,12\\ 
MeerKAT+        & 900$-$1670  & 770 &  1            & 80  & Fig.\,13\\
\hline
\end{tabular}
\label{frequency}
\end{table*}

\section{Low-frequency additional results}
\label{appendixB}

In this appendix, we explore the possibility to use low-frequency SKA-LOW data at higher spectral resolution or  
over a larger bandwidth. The setup with higher spectral resolution (Fig.\,B.1), i.e. 12\,kHz, does not show significant differences with respect to Fig.\,\ref{fig5}, indicating that no bandwidth depolarization is taking place or, if present, its contribution is negligible. 
Observations over a larger frequency band, i.e. 110--350\,MHz, 
are only slightly more powerful (Fig.\,B.2). Please, note that we use the same colour-bar ranges as in Fig.\,6 for easier comparison.

%\begin{figure*}
% \sidecaption
%\centering
%\includegraphics[width=11 cm]{FigB1.png}
%\caption{As in Fig.\,\ref{fig5} but for SKA-LOW observations in the frequency range 110-166\,MHz, with a spectral resolution of 12\,kHz and  smoothed to a resolution of 80$^{\prime\prime}$.}
%\label{figB1}
%\end{figure*}

%\begin{figure*}
% \sidecaption
%\centering
%\includegraphics[width=11 cm]{FigB2.png}
%\caption{As in Fig.\,\ref{fig5} but for SKA-LOW observations in the frequency range 110-350\,MHz, with a spectral resolution of 45\,kHz and  smoothed to a resolution of 80$^{\prime\prime}$.}
%\label{figB2}
%\end{figure*}

\end{appendix}
\end{document}